\documentclass[prb,aps,reprint,twocolumn,showpacs,tightenlines,amsmath,amssymb]{revtex4}
\usepackage{graphicx}    
\usepackage{amsmath}     
\usepackage{amssymb}  
\usepackage{mathrsfs}
\usepackage{bm}                                          
\usepackage{colordvi}         
\usepackage{epsfig}
\begin{document}

\title{Microscopic theory for the Doppler velocimetry of spin propagation 
in semiconductor quantum wells}
\author{M. Q. Weng}
\email{weng@ustc.edu.cn}
\author{M. W. Wu}
\email{mwwu@ustc.edu.cn.}
\affiliation{Hefei National Laboratory for Physical Sciences at Microscale and
  Department of Physics, University of Science and Technology of China, Hefei,
  Anhui, 230026, China} 
 \date{\today}

\begin{abstract} 
  We provide a microscopic
  theory for the Doppler velocimetry of 
  spin propagation in the presence of
  spatial inhomogeneity, driving electric field and the spin orbit
  coupling in semiconductor quantum wells in a wide range of
temperature regime based  on the kinetic spin Bloch equation. 
  It is analytically shown that under an applied electric field, the
  spin density wave gains a time-dependent  phase shift $\phi(t)$. 
  Without the spin-orbit coupling, the phase shift increases linearly with
  time and is equivalent to a normal Doppler shift in optical
  measurements. Due to the joint effect of spin-orbit coupling and the
  applied electric field, the phase shift 
behaves  differently at the
  early and the later stages. At the early stage, the phase shifts are
  the same with or without the spin-orbit coupling. While at the
  later stage, the phase shift deviates from the normal Doppler one
  when the spin-orbit coupling is present. 
  The crossover time from the early normal Doppler behavior to the
  anomalous one at the later stage 
  is inversely proportional to the spin 
  diffusion coefficient, wave vector
  of the spin density wave and the
  spin-orbit coupling strength. In the high temperature regime, the
  crossover time becomes large as a result of the decreased spin
  diffusion coefficient. 
  The analytical results capture all
  the quantitative features of the experimental results, while 
  the full numerical calculations agree quantitatively well with
  the experimental data obtained from the Doppler 
  velocimetry of spin propagation [Yang {\it et al.}, Nat. Phys. {\bf
    8}, 153 (2012)].  We further predict
  that the coherent spin precession, originally thought
  to be broken down at high temperature, is robust up to the room
  temperature for narrow quantum wells. We point out that one has to
  carry out the experiments longer to see the effect of the coherent
  spin precession at higher temperature due to the larger crossover time. 
\end{abstract}


\maketitle

\section{Introduction}

Understanding the spin transport phenomena is one of the most
important issues in the fast developing field of
spintronics\cite{awschalomlossSamarth,zuticrmp,Fabianbook, 
dyakonovbook_08,Wu201061,HandbookOfSpinTransport}
since it is crucial to the realization of the spintronic devices, such
as spin-field-effect transistor.\cite{datta:665,appelbaum:262501,
Appelbaum:447.295,HyunCheolKoo09182009,Wunderlich24122010}
In the proposed Datta-Das transistor,\cite{datta:665}
the ``on'' and ``off'' states, distinguished by 
a $\pi$-phase difference in the spin precession mode, 
are switched by the gate voltage which controls the
coherent spin precession (CSP) of the passing carriers via the Rashba
spin-orbit coupling (SOC)\cite{rashba:jetplett.39.78}
acting as an effective magnetic field. 
To implement such devices, it is required that one is able to 
control and maintain spin polarization over a long enough
distance, preferably at room temperature. 

Experimentally, the real space spin transport 
in semiconductor is studied by using magneto-optic
imaging,\cite{PhysRevLett.94.236601,beck_06}
or through conductance/current
modulation.\cite{HyunCheolKoo09182009,Wunderlich24122010}
An important development in the 
quantitative study of the spin transport is 
carried out by using spin transient grating 
spectroscopy.\cite{Yang2012,PhysRevLett.76.4793,nature.437.1330,
carter:136602,weber:076604,PhysRevB.82.155324}
In these experiments, a spin density 
wave (SDW) with initial spin polarization $S_z(x,0)=S_0\cos (qx)$ is
created by two 
orthogonal linear polarized light beams at time $t=0$, where $q$ is
the wave vector of the SDW and $x$ is the position. 
Under the influence of an 
applied in-plane electric field and the SOC, SDW picks up a
phase and evolves into 
$S_z(x,t)=S^0_z(q,t)\cos (qx-\phi(t))$.
By optically monitoring the temporal evolution of 
the amplitude $S^0_z(q,t)$, one obtains 
the spin diffusion coefficient and relaxation
rate.\cite{PhysRevLett.76.4793,nature.437.1330,
carter:136602,weber:076604,PhysRevB.82.155324,Yang2012,weng:063714}
Very recently, the spin drifting and CSP were studied by
the Doppler velocimetry which monitors the phase 
shift $\phi(t)$.\cite{PhysRevLett.106.247401,Yang2012}
For pure spin drift without spin precession, the phase shift is simply
$\phi(t)=qv_dt$, where $v_d$ stands for the drift velocity. The
linear increase of phase shift with time is equivalent to a Doppler
shift $\Delta\omega=v_dq$. 
When both the CSP and drifting are present,
the phase shift deviates from this simple relation and behaves
anomalously. In Ref.~[\onlinecite{Yang2012}], Yang {\it et al.} reported
that at low temperature ($T=30$~K), $\phi(t)$ deviates strongly from the
simple relation of $qv_dt$ and clearly shows the anomalous behavior
caused by the CSP. However, once the temperature rises to 150~K, the
anomalous behavior of the phase shift disappears in the time frame of
the observation. It was then concluded that CSP breaks down at
high temperature, although the mechanism of the disappearance of CSP
is not clear.\cite{Yang2012}
The disappearance of CSP at high temperature was claimed to be in
consistence with the previous experimental results in the prototypes
of Datta-Das transistors.\cite{HyunCheolKoo09182009}

The temporal evolution of the SDW contains all of the important 
information of spin transport, such as spin diffusion coefficient,
spin mobility and CSP. The transient spin grating spectroscopy
together with the Doppler velocimetry  therefore enable one to
quantitatively study the spin transport in semiconductors. 
However, the dynamics of SDW is quite 
complex when spatial inhomogeneity, applied electric field as
well as the SOC are present. To correctly extract the information from the
experimental data, a thorough understanding of spin transport is
in demand. Without the electric field, the amplitude $S^0_z(q,t)$
decays biexponentially when the SDW diffuses along the 
$[1\bar{1}0]$ crystal
axis in (001) GaAs quantum wells (QWs) as spin rotates along the net
effective magnetic field due to the SOC and
diffusion.\cite{nature.437.1330,weng:063714,PhysRevB.84.075326}
Based on the kinetic spin Bloch equation 
(KSBE) approach,\cite{PhysRevB.66.235109,Wu201061}
it is shown that the information
about spin diffusion coefficient, CSP and spin relaxation can be extracted
from the wave-vector dependence of the two decay
rates.\cite{weng:063714} In this paper we will further extend the
theory to 
include the electric field and 
show, both analytically and numerically, that the theoretical and
experimental results agree well with each other. We will 
demonstrate that the CSP is robust even in high temperature regime
for narrow QWs and point out that to observe the effect of
the CSP at high temperature, one has to carry out the observation for
a longer time than the case at low temperature. 

\section{Analytical results of the evolution of the SDW}

As will be shown later, the SDW transporting along 
the $[1\bar{1}0]$ crystal axis with the wave vector $q$
in a $(001)$ GaAs QW evolves as
\begin{eqnarray}
 S_z(x,t)&=&S_z(q,0)\exp[-(Dq^2+1/\tau_s)t]/2 \nonumber \\
&& \mbox{ }\times \Bigl\{e^{-2Dqq_0t}\cos[qx-v_d(q+q'_0)t]
\nonumber \\ &&  \mbox{ }
+  e^{2Dqq_0t}\cos[qx-v_d(q-q'_0)t]
  \Bigr\} \;,
  \label{eq:SDW_t}
\end{eqnarray}
when the SOC is weak enough. Here 
$D$, $\tau_s$ and $v_d$ are 
the spin diffusion coefficient, spin
relaxation time of the SDW and
the drift velocity under the electric field
$E$, respectively. $q_0=m^{\ast}(\hat{\beta}+\alpha)$ and 
$q'_0=m^{\ast}(\hat{\beta}'+\alpha)$
with 
$m^{\ast}$ representing the effective mass, 
$\alpha$ being the Rashba coefficient\cite{rashba:jetplett.39.78} and 
$\hat{\beta}$, $\hat{\beta}'$ both standing for the coefficients of the 
linear Dresselhaus term\cite{dp} 
with corrections from the cubic Dresselhaus terms.\cite{weng:063714}
Without the applied electric field, $v_d=0$ and the amplitude of the SDW
decays biexponentially with fast and slow rates $Dq^2+1/\tau_s\pm
2Dqq_0$.\cite{nature.437.1330,weng:063714}
When there is an applied electric field but without the CSP
($q_0=q_0'=0$), the SDW decays exponentially and gains a 
phase shift which changes linearly with time with a slope 
$qv_d$, which 
is equivalent to a normal Doppler shift in experiments.
With the CSP, the situation is more complex. 
For small time $t\ll 1/(4D|qq_0|)$, the fast and the slow modes 
share the same weights. Therefore the phase shift of the SDW reads
\begin{equation}
  \label{eq:phase_smallt}
  \phi(t)\simeq [v_d(q+q_0')t+v_d(q-q_0')t]/2
  = v_dqt \;.
\end{equation}
Nevertheless,  for large time $t\gg 1/(4D|qq_0|)$, the slow mode
dominates and the phase shift becomes  
\begin{equation}
  \label{eq:phase_larget}
  \phi(t)\simeq v_d(q-q'_0)t \;.
\end{equation}
That is, in the presence of the CSP, the phase shift first changes linearly
with a slope $qv_d$, same as the normal Doppler one without the CSP. 
After some time, the slope deviates from the normal one and 
reduces to $v_d(q-q'_0)$.
The sign of the slope reverses when $q<q'_0$. In the 
special case of $q=q_0'$, the phase approaches a stationary value at
large time. 
The crossover time of $\phi(t)$ from the normal Doppler behavior in
the early 
stage to the anomalous one at the later stage is
about  $t_c\sim 1/(4D|qq_0|)$. In the case of small 
diffusion coefficient, wave vector of SDW or the SOC, this crossover time
can be larger. 

Equation~(\ref{eq:SDW_t})
captures all of the qualitative features of the
experimental results\cite{Yang2012}
at low temperature, from the two modes in the
temporal evolution of SDW when $E=0$ (hence $v_d=0$), to the details of
how the phase 
changes with time and wave vector when $E\not=0$. Specifically, the
temporal evolution of the phase can be divided into two stages. In the
early stage, the phase increases with time with a steeper slope. The
slope becomes flatter in the later stage, or even reverses its sign
for small wave vectors. The larger the wave vector is, the quicker the
phase behavior changes from the early stage to the later stage.
All these features qualitatively agree with Eq.~(\ref{eq:SDW_t}).

As a result of the increasing electron-phonon scattering and spin
Coulomb drag,\cite{amico_00,nature.437.1330,takahashi_08,badalyan_08}
the spin diffusion coefficient $D$
decreases with the increase of temperature.
It is therefore expected that at high temperature the crossover time
is larger than that at low temperature. 
With the correction of the cubic Dresselhaus term, $q_0$ also 
reduces as the temperature rises and the crossover time is 
further prolonged. If the contribution from the cubic Dresselhaus term
is so large that $q_0$ approaches zero, then the CSP breaks down completely
at high temperature. 
In the recent experiments, the CSP is thought to be broken down at high
temperature based on the lack of the clear anomalous behavior in the
phase shift from the observation in a limited time regime. 
The cubic Dresselhaus term is speculated to be 
the main cause for the breaking
down.\cite{Yang2012} However, this is highly unlikely because 
$q_0$ at low temperature only differs by a few 
percents from the one at high temperature
for the narrow QW used in the experiments.
It should be pointed out that, since the behaviors of phase in the
early stage are the same with or without the CSP, one should be cautious
in determining the existence of such CSP from the experimental data in
limited time regime, especially when the crossover time $t_c$ is
large. In our opinion, purely from the existing experiments, it is
inconclusive to determine whether the CSP survives or not at high
temperature. 

\begin{figure}[htbp]
  \centering
  \epsfig{file=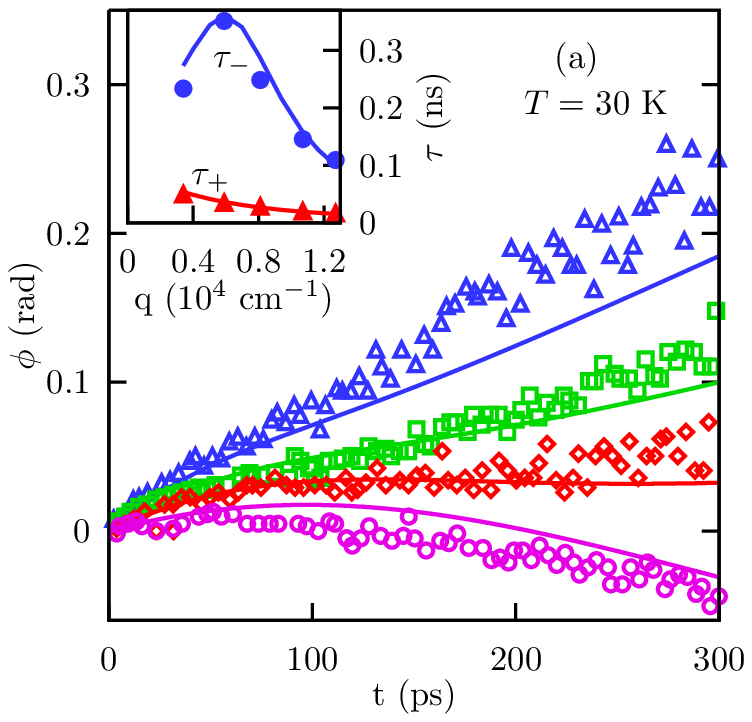,width=0.85\columnwidth}
  \epsfig{file=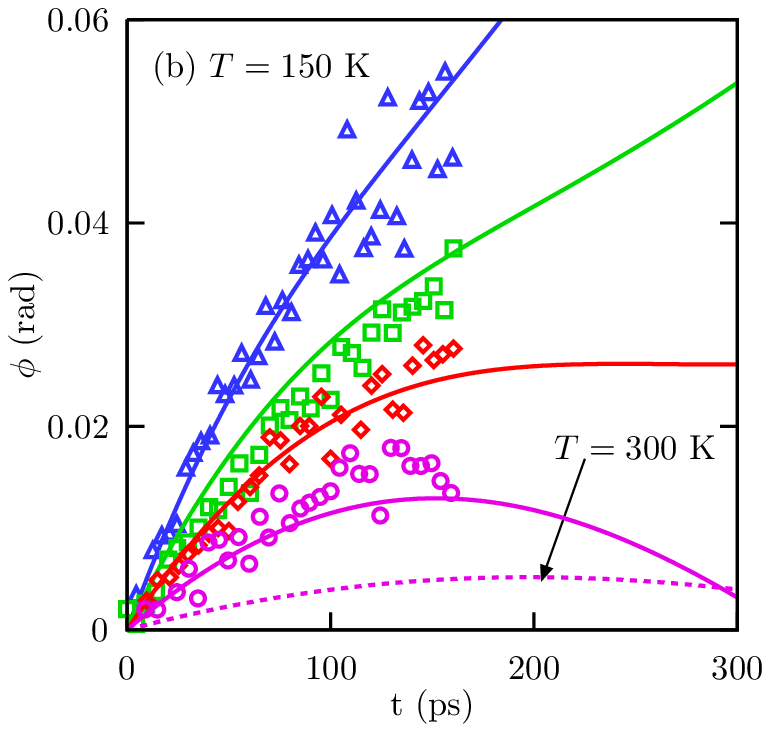,width=0.85\columnwidth}
  \caption{(Color online) 
    Phase shifts of drifting spin grating under an applied electric
    field of $E=2$~V/cm  for different wave vector at (a) $T=30$~K and
    (b) $T=150$~K. 
    Blue Curve/Triangle: $q=1.07$~cm$^{-1}$; Green Curve/Square: 
$q=0.81$~cm$^{-1}$; 
    Red Curve/Diamond: $q=0.59$~cm$^{-1}$; Purple Curve/Circle:
    $q=0.34$~cm$^{-1}$, respectively.  
    Inset of (a): Fitting of spin relaxation times $\tau_+$ 
    (Red Curve/Circle) and $\tau_-$ (Blue Curve/Triangle) at
    $T=30$~K. 
    The dashed purple curve in (b) is the phase shift for
    $q=0.34$~cm$^{-1}$ at $T=300$~K. 
    All the curves in the figures are from theoretical calculation whereas 
    the symbols are the experimental data from Ref.~[\onlinecite{Yang2012}]. 
  }
  \label{fig:phase}
\end{figure}  

To derive the solution [Eq.~(\ref{eq:SDW_t})] and to 
determine if the CSP is stable at high temperature, we turn
to the full KSBEs for the spin transport in a (001) GaAs
QW grown along the
$z$-axis\cite{Wu201061,HandbookOfSpinTransport,PhysRevB.66.235109} 
\begin{eqnarray}
&&{\partial\rho_{\mathbf{k}}(x,t) \over \partial t} = -
    eE(x){\partial\rho_{\mathbf{k}}(x,t)\over\partial k_x}+
    {k_x\over m^{\ast}}
    {\partial\rho_{\mathbf{k}}(x,t)\over\partial x} \nonumber 
\\ &&\mbox{ }
    + i[
    \mathbf{h}_{tot}(\mathbf{k}))\cdot
    \mbox{\boldmath $\sigma$}/2, 
\rho_{\mathbf{k}}(x,t)]
+\left.{\partial \rho_{\mathbf{k}}(x,t)\over
  \partial t}\right|_{\mathtt{s}}.
\label{eq:kinetic}
\end{eqnarray}
Here we assume that the transport direction is along the
$x$-axis. $\rho_{\mathbf{k}}(x,t)$ are the  density matrices
of electron with momentum
$\mathbf{k}=(k_x,k_y)=(k\cos\phi, k\sin\phi)$ at position $x$. 
The right hand side of 
Eq.~(\ref{eq:kinetic}) describes 
the drift of electrons
driven by the electric field $E(x)$, 
diffusion caused by the spatial inhomogeneity,
spin precession
around the total magnetic field ${\bf h}_{tot}({\bf k})$
and all the scattering, respectively.
The total magnetic field is composed of the external
magnetic field $\mathbf{B}$ in the 
Voigt  configuration, the effective
magnetic field $\mathbf{h}(\mathbf{k})$ due to the 
SOC as well as the one from the Hartree-Fock term of the electron-electron
Coulomb interaction. The expressions for the Hartree-Fock and
the scattering terms are given in detail in Refs.\
[\onlinecite{PhysRevB.69.245320,Wu201061}].
$\mathbf{h}(\mathbf{k})$ contains the Dresselhaus and the Rashba
terms:\cite{dp,rashba:jetplett.39.78}
\begin{eqnarray}
\mathbf{h}(\mathbf{k})&=&\beta
(-k_x\cos 2\theta+k_y\sin 2\theta,
k_x\sin 2\theta+k_y\cos 2\theta,0)
\nonumber\\&&
+\gamma({k_x^2-k_y^2\over 2}\sin 2\theta
+k_xk_y\cos 2\theta)(k_y,-k_x,0)
\nonumber\\&&
+\alpha(k_y,-k_x,0)\ ,
\label{eq:hk}
\end{eqnarray}
where $\theta$ is the angle between $x$-axis (the spin
injection/diffusion direction) and the 
$(100)$ crystal axis.\cite{cheng:205328,weng:063714} 
$\beta=\gamma \pi^2/a^2$ with
$\gamma$ being the Dresselhaus coefficient.\cite{dp}
$\alpha$ represents the Rashba parameter which depends on the electric
field along the  growth direction of the QW.
Note that we have included the corrections
from the cubic Dresselhaus term.

By expanding the density matrix 
$\rho_{\bf k}(x,t)=\sum_l\rho_l(x,k,t)e^{il\phi}$, the KSBEs can be 
written as a series of 
coupled equations for $\rho_l(x,k,t)$. 
By neglecting the Hatree-Fock term \footnote{For systems with small
  spin polarization, Hatree-Fock term can be neglected.
  \cite{Wu201061}}
and the inelastic scattering and 
using the fact that the spin density is 
${\bf S}(x,t)=\sum_{\bf k}\mathtt{Tr}\{
\mbox{\boldmath $\sigma$\unboldmath}
\rho_{\bf k}(x,t)\}
=\int_0^{\infty} \mathtt{Tr}[
\mbox{\boldmath $\sigma$\unboldmath}
\rho_0(x,k,t)]
kdk/2\pi$, one finds that, to the leading order, 
under a uniform applied electric field $E$ the spin density obeys the
following equation
\begin{equation}
  {\partial {\bf S}\over \partial t}
    =
    D {\partial ^2{\bf S}\over\partial x^2}
    +v_d{\partial {\bf S}\over\partial x}
    -\bm{\mathscr{R}}\cdot
    {\bf S}
    +2D\bar{{\bf h}}\times {\partial {\bf S}\over \partial x}
    +v_d\bar{{\bf h}}^\prime\times {\bf S}, 
\label{eq:S_env}    
\end{equation}
in which the diffusion coefficient is
$D=\langle {k^2\tau_1/(2m^{\ast 2})}\rangle$ and
the drift velocity reads
$v_d=\langle {eE\tau_1/m^{\ast}}\rangle$. 
The first three terms of the right hand side of Eq.~(\ref{eq:S_env})
describe the diffusion caused by
the spatial inhomogeneity,
drift driven by the electric field as well as the relaxation
of the spin polarization, 
respectively. 
The fourth term stands for the spin precession around the net effective
magnetic field (propotional to $\bar{\bf{h}}$) due to the joint effect
of the SOC and the diffusion, whereas
the last term is the precession around another net
effective magnetic field 
(propotional to $\bar{\bf{h}}^\prime$) due to the joint effect
of the SOC and the drift, with 
\begin{equation}
  \label{eq:h_diff}
  \bar{\mathbf{h}}
  =m^{\ast}(-\hat{\beta} \cos 2\theta, \hat{\beta}\sin 2\theta-\alpha, 0)
\end{equation}
and 
\begin{equation}
  \label{eq:h_drift}
  \bar{{\bf h}}^\prime=
  m^{\ast}(-\hat{\beta}^\prime \cos 2\theta, \hat{\beta}^\prime
\sin 2\theta-\alpha, 0).
\end{equation}
Here 
$\hat{\beta}=\beta-\gamma \langle k^2\rangle/4$ and 
$\hat{\beta}^\prime=\beta-\gamma \langle k^2\rangle/2$.

As noted in Ref.~[\onlinecite{weng:063714}], the solution to this
equation is quite complex in general situation. For transport
along the $[110]$ or $[1\bar{1}0]$ crystal axes, or in the case that only the
Dresselhaus or Rashba term is important, the solution is simpler.
To understand the existing experimental results, here
we focus on the transport along the $[1\bar{1}0]$ crystal axes ($x$-axis).
In this case, 
$\theta=-\pi/4$, 
Eqs.~(\ref{eq:h_diff}) and (\ref{eq:h_drift})
can be further simplified as
$\bar{{\bf h}}=-q_0{\bf e}_y$ and $\bar{{\bf h}}^\prime=-{q}_0^\prime{\bf
  e}_y$ with ${\bf e}_y$ being the unit vector along the $y$-axis 
([110] crystal axis), 
$q_0=m^{\ast}(\hat{\beta}+\alpha)$ and
$q^\prime_0=m^{\ast}(\hat{\beta}^\prime+\alpha)$. 
Furthermore,
the relaxation matrix
$\bm{\mathscr{R}}=\mathtt{diag}\{1/\tau_x,1/\tau_y,1/\tau_z\}$ is
diagonal, with $1/\tau_{x}=\langle(\hat{\beta}+\alpha)^2
k^2\tau_1/2\rangle+\langle \gamma^2k^6\tau_3/32\rangle$,  
$1/\tau_{y}=\langle(\hat{\beta}-\alpha)^2
k^2\tau_1/2\rangle+\langle \gamma^2k^6\tau_3/32\rangle$ and 
$1/\tau_z=1/\tau_x+1/\tau_y$ being spin relaxation rates of spin
components along the $x$, $y$ and $z$ direction, respectively. 
$1/\tau_l=\int_0^{2\pi}{1\over\tau(k,\theta)}\cos
(l\theta)d\theta/2\pi$ with $\tau(k,\theta)$ standing for the momentum
relaxation time due to the electron-impurity scattering. 
For a system near the equilibrium, 
$\langle\cdots\rangle=\int \cdots \partial
  f(\varepsilon_{\mathbf{k}})/\partial \varepsilon_{\mathbf{k}}
d^2\mathbf{k} /\int \partial
  f(\varepsilon_{\mathbf{k}})/\partial \varepsilon_{\mathbf{k}}
d^2\mathbf{k}$ with $f(\varepsilon)$ being the Fermi  distribution
function. 

The right hand side of Eq.~(\ref{eq:S_env}) describes the spin 
diffusion, drifting, and spin precession around the net effective
magnetic field as well as  the spin relaxation.
It is noted that a similar equation is derived by 
linear response theory,\cite{PhysRevB.70.155308}
kinetic theory\cite{halperin_04,bleibaum_06,
bryksin:165313,
hruska:075306,PhysRevLett.94.236601}
random walk model\cite{PhysRevB.82.155324} and 
Monte Carlo simulation.\cite{PhysRevB.77.045323}
Comparing to these approaches, instead of using phenomenological
parameters, we obtain all the transport parameters fully microscopically.
 Moreover, we also correctly take the correction from the
cubic Dresselhaus term into account. 

For initially $z$ polarized SDW with wave vector $q$, the solution 
to Eq.~(\ref{eq:S_env}) reads 
\begin{equation}
  \label{eq:Szt}
  S_z(q,t)=S_z(q,0)
  [\lambda_+(q)e^{-\Gamma_+(q)t}
  +\lambda_-(q)e^{-\Gamma_-(q)t}], 
\end{equation}
with $S_z(q,t)$  being the Fourier component of SDW and 
$S_z(q,0)$ standing for the initial spin density, 
\begin{equation}
  \label{eq:Gamma}
  \Gamma_{\pm}(q)=Dq^2-iv_dq+{1/\tau_s}\pm 
  {\Delta/2\tau_y}, 
\end{equation}
$    \lambda_{\pm}(q)=\bigl[1\pm
    {1/\Delta}\bigr]/2 , 
$
where 
$1/\tau_s=(1/\tau_x+1/\tau_z)/2$ 
and
$\Delta=\sqrt{1+4\tau_y^2(2Dqq_0-iv_dq_0')^2}$.
The result is similar to the one without the applied electric field:
the temporal evolution of $S_z(q,t)$ is composed of two modes, 
with decay rates being 
$\Re\{\Gamma_+(q)\}$ respectively.\cite{weng:063714}
For small electric field, 
the difference between the decay rates is
quadratic in the field. Therefore, the
electric field has only marginal effect on the decay rates. 
However, the electric field introduces 
additional spin precession, causing 
$S_z(q,t)$ to oscillate with frequency 
$|\Im\{\Gamma_{\pm}(q)\}|$ which is linear to the electric field for
small field. 
In the case with $|\Delta|\gg 1$, such as in the system with 
weak SOC or 
in the special case when $\beta\simeq \alpha$
(in these cases $\tau_y$ becomes very large), 
the solution can be further
simplified to 
$\lambda_{\pm}\simeq 1/2$ and 
\begin{equation}
  \label{eq:Gamma_largetauy}
  \Gamma_{\pm}(q)=Dq^2\pm 2Dqq_0-iv_d(q\pm q'_0)+{1/\tau_s}, 
\end{equation}
one then gets time evolution of the SDW as Eq.~(\ref{eq:SDW_t}). 
It is noted that 
by using their dependencies on the SOC and momentum relaxation time,
one can prove that
$1/\tau_s=(1/\tau_x+1/\tau_z)/2=
1/\tau_x+1/(2\tau_y)
\ge 1/\tau_x\ge
\langle [m^{\ast}(\hat{\beta}+\alpha)]^2k^2\tau_1/2m^{\ast 2}
\rangle\simeq 
\langle k^2\tau_1/2m^{\ast 2}\rangle
\langle [m^{\ast}(\hat{\beta}+\alpha)]^2\rangle
=Dq_0^2$.
Therefore the real parts of the decay rates 
$\Re\{\Gamma_{\pm}(q)\}=Dq^2\pm 2Dqq_0+1/\tau_s\ge Dq^2\pm 2Dqq_0+Dq_0^2
=D(q\pm q_0)^2$ are always non-negative. This 
indicates that the amplitude of the SDW does not increase with time.
For the special SDW with $q=q_0$, $\Gamma_-(q)$ is pure
imaginary, corresponding to the so call persistent spin
helix\cite{PhysRevB.70.155308,halperin_04,weber:076604}
when $\alpha=\beta$ and the cubic Dresselhaus terms are neglected.

\section{Numerical solution to the full KSBE's for the
SDW}

To clarify whether the CSP survives at high temperature, 
one has to calculate the temporal evolution of the
SDW by numerically solving the {\em full} KSBEs.
In the calculation, we include all the 
relevant scattering, such as the electron-impurity, electron-AC
phonon, electron-LO phonon as well as electron-electron 
Coulomb scattering.\footnote{In our calculation, 
we use dynamically screened Coulomb potential with 
the random phase appriximation. To compensate the local field corrections, we 
use the exact two-dimensional Coulomb interaction to 
replace the quasi-two-dimensional one at 30~K. This replacement is
known to be a good approximation for QW with width about
10~nm below 100~K.\cite{badalyan_08,nature.437.1330}}
The material and structural parameters are chosen from
the available experimental data directly or by fitting the
corresponding experimental data: QW width
$a=9$~nm, and the electron density $N_e=1.9\times
10^{11}$~cm$^{-2}$. 
The effective impurity density $N_i$ is set to be $0.12 N_e$ by 
fitting the mobility under the laser intensity 
of 0.25~$\mu$J$\cdot$cm$^{-2}$.
\footnote{
  Without the laser pumping, the impurity density is $0.02 N_e$
  by fitting the electron mobility at 5~K. Using this set of data, the
  theoretical mobility and the experimental results agree with each
  other within 15\% margin for the whole experimental temperature
  regime. Under the laser pumping, electrons suffer additional
  scattering from the excited holes, which are replaced by additional
  impurities in our calculation for simplification. 
  It is found that with an effective impurity density of $0.12 N_e$,
  one recovers the experimental mobility data at $T=30$~K under the
  laser intensity of 0.25~$\mu$J$\cdot$cm$^{-2}$. }
The Rashba coefficient $\alpha=0$ since
the QW is symmetric. The Dresselhaus coefficient
$\gamma$ is set to be $12$~meV$\cdot$\AA$^{-3}$ by fitting the spin
relaxation time at $T=30$~K. 
The KSBEs used here are valid unless the higher subbands of QW are
significantly occupied by electrons. For the parameters used here,
this does not  
happen until $T\sim 800$~K, well above the room temperature. 
For wave vector-dependent spin
relaxation times at $T=30$~K, shown in the inset of
Fig.~\ref{fig:phase}(a), the theoretical and experimental results are
also in very good agreement. To see the effect of the CSP, we plot 
the phase shifts $\phi(t)$ of the SDW 
under the influence of an applied electric field of
$E=2$~V/cm at different gratings 
as function of time in Fig.~\ref{fig:phase} for
$T=30$ and $150$~K. For comparison, we also plot
the experimental data in the figure. It can be seen that
theoretical and experimental results agree well with each other. 
More importantly, at $T=150$~K both
theoretical and experimental results show influence of the CSP on
$\phi$ near $t=150$~ps, where the slopes of $\phi$
become flat. In the theoretical 
calculation, the effect is revealed more clearly for SDW with 
$q=0.34\times 10^{4}$~cm$^{-2}$ after about 200~ps when the crossover
from positive slope to the negative one is completed. 
The calculations at room temperature is also carried out and CSP is
found to be robust 
even at room temperature. But the crossover from the early 
to later stage is further delayed, {e.g.}, for $q=0.34\times
10^4$~cm$^{-1}$ the 
crossover from positive slope to the negative one finishs at around
250~ps, as shown in Fig.~\ref{fig:phase}(b). 
From these calculations one concludes that CSP indeed survives
at room temperature for the narrow QWs studied here, but one has to
carry out the observation for a longer time to observe its effect on
the phase experimentally. 
This result is in consistence with our previous study on spin
transport in GaAs QWs\cite{PhysRevB.66.235109} and
the Datta-Das transistor,\cite{0268-1242-26-7-075005}
in which it is shown that the high temperature spin precession 
still exists, although the amplitude is much
weaker than those at low temperatures.

\section{Conclusion}
In conclusion, we provide a microscopic theory for the Doppler
velocimetry of spin propagation 
in the presence of spatial inhomogeneity, driving 
electric field and the CSP due to the joint effect of the SOC and
transport in a wide range of temperature regime. 
Applying this theory to study the transport of SDW, 
we analytically show that in the presence of the electric
field the SDW gains a time-dependent phase shift $\phi(t)$. 
Without the CSP, $\phi(t)$ grows linearly with time with slope $qv_d$ for
the SDW with wave vector
$q$, which is equivalent to a normal Doppler
shift in optical measurements. 
Due to the CSP caused by the net effective magnetic field from the joint
effect of the SOC and transport, the short time and the long time
behaviors of phase shift are different.  
At the early stage
$\phi(t)$ grows with time in the same
way as that of without the CSP, {\em i.e.} with 
a normal Doppler slope $qv_d$. 
At the later stage the slope reduces to $(q-q_0)v_d$, deviating from
the normal Doppler one. For
small $q$ the slope at the later stage even reverses its sign. 
The crossover time from the early to the later
stages is inversely proportional to the spin diffusion coefficient $D$,
the wave vector
of the SDW $q$ and the SOC strength. Since $D$ decreases
with the increase of temperature, the crossover from the early
to the later stage at high temperature would be prolonged. 
Our numerical calculations, which include all the relevant scattering
such as the electron-impurity, electron-AC phonon, electron-LO
phonon and electron-electron Coulomb scattering, agree 
quantitatively well with the
existing experimental results. By extending the calculation time beyond
the experiment regime, we predict 
that the CSP, originally thought to be broken at high temperature, is
robust and stable up to the room temperature. We further point out  
that to observe the effect of the CSP on the phase shift a longer
measurement time is required at higher temperature.

\begin{acknowledgments}

We would like to thank J. Orenstein and L. Y. Yang for
  providing details of their experiments. 
This work was supported by the National Basic Research Program of China under
Grant No.~2012CB922002 and the Strategic Priority Research Program of
the Chinese Academy of Sciences under Grant No. XDB01000000. 
MQW was also supported by the 
 Anhui Natural Science Foundation under Grant
No.~11040606Q46. 
\end{acknowledgments}


\end{document}